\documentclass[prb]{revtex4}
\usepackage{psfig,epsfig}
\usepackage{amsmath,amsfonts,amssymb}
\begin{document}

\title{From Slow to Superluminal Propagation: Dispersive Properties of
  Surface Plasmon Polaritons in Linear Chains of Metallic
  Nanospheroids}

\author{Alexander A. Govyadinov$^a$ and Vadim A. Markel$^{a,b}$}
\affiliation{Departments of $^a$Bioengineering and $^b$Radiology, University
  of Pennsylvania, Philadelphia, PA 19104}

\begin{abstract}
  We consider propagation of surface plasmon polaritons (SPPs) in
  linear periodic chains (LPCs) of prolate and oblate metallic
  spheroids. We show that the SPP group velocity can be efficiently
  controlled by varying the aspect ratio of the spheroids. For
  sufficiently small aspect ratios, a gap appears in the first
  Brillouin zone of the chain lattice in which propagating modes do
  not exist. Depending on the SPP polarization, the gap extends to
  certain intervals of the Bloch wave number $q$. Thus, for transverse
  polarization, no propagating SPPs exist with wave numbers $q$ such
  that $q_c^\perp < \vert q \vert < \pi/h$, $h$ being the chain
  period. For longitudinally polarized SPPs, the gap spans the
  interval $\vert q \vert <q_c^\parallel$. Here $q_c^\perp$ and
  $q_c^\parallel$ are different constants which depend on the chain
  parameters, spheroid aspect ratio and its orientation with respect
  to the chain axis. The dependence of the dispersion curves on the
  spheroid aspect ratio leads to a number of interesting effects. In
  particular, bandwidth of SPPs that can propagate in an LPC can be
  substantially increased by utilizing prolate or oblate spheroids.
  When $q$ is close to a critical value, so that $\vert q -
  q_c^\perp\vert \ll \pi/h$ or $\vert q - q_c^\parallel\vert \ll
  \pi/h$, the decay length of the SPPs is dramatically increased. In
  addition, the dispersion curves acquire a very large positive or
  negative slope. This can be used to achieve superluminal group
  velocity for realistic chain parameters. We demonstrate superluminal
  propagation of Gaussian wave packets in numerical simulations. Both
  theory and simulations are based on Maxwell equations with account
  of retardation and, therefore, are fully relativistic.
\end{abstract}

\date{\today}
\maketitle

\section{Introduction}
\label{sec:intro}

Propagation of surface plasmon polaritons (SPPs) in linear periodic
chains (LPC) of metal nanoparticle has been in the focus of
considerable recent
attention~\cite{weber_04_1,simovski_05_1,koenderink_06_1,fung_07_1,park_04_1,citrin_05_1,citrin_06_1,markel_07_2}.
The interest is, in part, motivated by the potential application of
such chains as sub-wavelength plasmonic
waveguides~\cite{quinten_98_1,brongersma_00_1,maier_03_1}. Since
energy in a waveguide can be transported in the form of wave packets,
the dispersion relation becomes of primary importance.  In the case of
LPCs, the dispersion relation is a mathematical dependence of the SPP
frequency $\omega$ on its Bloch wave number, $q$. It is important to
emphasize that the elementary excitations (electromagnetic modes) in
LPCs are Bloch waves rather than sinusoidal waves which propagate in
continuous media. Only when the Bloch wave number is sufficiently
small, so that $qh\ll 1$, where $h$ is the chain period, can we
neglect the mathematical distinction between the Bloch and the
sinusoidal waves and treat an LPC as an essentially continuous system.
How strong the above inequality should be is not evident {\em a
  priori}; we will comment on this question in the concluding part of
the paper.

Dispersion curves have been previously computed for polarization waves
propagating in periodic chains of Drudean spherical
nanoparticles~\cite{weber_04_1,simovski_05_1,koenderink_06_1,fung_07_1}.
It was found that there exist excitations with frequencies $\omega$
and Bloch wave numbers $q$ such that $\omega < qc_h$, $c_h$ being the
speed of light in the surrounding (host) medium. The phase velocity of
such excitations, $v_p = \omega/q$, is less in magnitude than $c_h$.
SPPs with $\vert v_p\vert < c_h$ are considered to be outside of the
``light cone'' and can propagate along the chain without radiative
losses, similarly to plane waves in homogeneous dielectrics.  We will
reserve the term ``SPP'' specifically for this type of excitations. In
a chain of perfectly conducting (lossless) particles, an SPP can
propagate infinitely without decay. Of course, absorptive (Ohmic)
losses in realistic metals always result in exponential spatial decay
of SPPs.

The property of the dispersion curve $\omega(q)$ which is specific to
the case of LPCs made of spherical particles is that it is very
flat~\cite{weber_04_1,simovski_05_1,park_04_1}, with only a weak
dependence of the frequency on the Bloch wave number. The dispersion
curves are particularly flat for SPPs polarized transversely to the
chain. This results in very small group velocities, $v_g \ll c_h$. A
factor $v_g/ c_h \sim 10^{-2}$ is typical. One practically important
consequence of the dispersion curve flatness is a relatively narrow
SPP bandwidth. That is, an SPP can be excited in a chain by a
spatially-localized external source only in a narrow band of
frequencies. This can be expected to significantly limit the potential
application of LPCs as optical waveguides.

Maier {\em et al.} have pointed out that the use of spheroidal rather
than of spherical nanoparticles can result in an increased SPP
bandwidth and, correspondingly, in a longer propagation
distance~\cite{maier_02_1}. In the above work, the bandwidth was
defined as twice the spectral shift of the nearly homogeneous SPP
(characterized by $q=0$) with respect to the plasmon peak of an
isolated nanoparticle and it was assumed that the propagation
distance is proportional to the inverse of the latter. The results
were confirmed by FDTD simulations in an LPC of seven prolate
nanospheroid whose longer axis was perpendicular to the chain.

In this paper, we further investigate the effects of nonsphericity of
the LPC constituents. We compute the dispersion relations in such LPCs
in the dipole approximation. We find that the dispersion curves in
LPCs are dramatically altered by replacing spherical particles with
prolate or oblate spheroids. In particular, the SPP bandwidth can be
significantly increased, in agreement with the results of Maier {\em
  et al.}. Here, however, we define the bandwidth as the range of
frequencies in which efficient SPP transport along the chain is
possible. This includes modes with various values of $q$, including
those with $q\sim \pi/h$. We also find that the use of oblate
spheroids (nanodisks) whose shorter semiaxis is parallel to the chain
is even more beneficial as it allows to achieve the desired effect at
relatively modest values of the aspect ratio. The increased bandwidth
is expected to result in a higher maximum bit rate and longer
propagation distances for signals transported by an LPC waveguide.

We further report that at some critical values of the spheroid aspect
ratio, gaps appear in the first Brillouin zone of the lattice.
Propagating SPPs do not exist when the Bloch wave number $q$ is inside
one of such gaps. When $q$ is near the gap edge, a number of
interesting phenomena take place.  First, the propagation distance
(the decay length) is dramatically increased, as compared to the same
quantity when $q$ is far from the edge.  Second, the dispersion curves
acquire very large positive or negative slopes.  In this case,
relatively fast ($c_h - \vert v_g \vert \ll c_h$) and even
superluminal ($\vert v_g \vert > c_h$) wave packet propagation can be
obtained.  Note that superluminal group velocity does not contradict
special relativity~\cite{bolda_94_1}.  Superluminal wave packets exist
in nature and were observed
experimentally~\cite{wang_00_1,gehring_06_1}.

Theory and numerical simulations presented below are based on the
dipole approximation. The use of this approximation dictates that the
inter-particle spacings are larger than a certain threshold at which
excitation of higher multipole moments in nanospheroids becomes
non-negligible. This has limited the range of chain parameters
considered in this paper. We, however, expect on physical grounds that
using chains with smaller inter-particle separations may be
beneficial. For instance, we expect that the SPP propagation length in
such chains can be increased. However, in order to obtain quantitative
results in that limit, a considerably more complex mathematical
formalism must be used. The latter has been developed by Park and
Stroud~\cite{park_04_1} for chains of spherical nanoparticles in the
quasistatic limit. Unfortunately, generalization of this formalism to
non-spherical particles and beyond the quasistatic approximation
(which we deem to be essential for describing SPP propagation in long
chains, as was confirmed recently in the experiments by Koenderink
{\em et al.}~\cite{koenderink_07_1}) appears to be problematic.

The paper is organized as follows. In the Section~\ref{sec:model}, we
describe and justify the basic model used to simulate Bloch waves and
wave packets in LPCs. In Section~\ref{sec:dispersion}, we compute the
dispersion curves for SPPs in chains of prolate and oblate
nanospheroids. In Section~\ref{sec:l}, we discuss the attenuation of
SPPs due to Ohmic losses in LPCs and, for comparison, in metallic
nanowires. In Section~\ref{sec:packets}, we describe direct numerical
simulations of wave packet propagation in LPCs. Section~\ref{sec:disc}
contains a summary and a discussion of obtained results.

\section{The Basic Model}
\label{sec:model}

Consider a linear periodic chain of identical metallic spheroids with
semiaxes $a$ and $b$ ($a \geq b$). We will discuss below two different
cases. In the first case, the chain is made of prolate spheroids whose
axis of symmetry (which coincides with the longer axis) is
perpendicular to the chain. In the second case, the chain is made of
oblate spheroids whose axis of symmetry (which coincides with the
shorter axis) is parallel to the chain. In both cases, the longer axes
of the spheroids are perpendicular to the chain and the eccentricity,
$e$, is given by

\begin{equation}
\label{ecc_def}
e = \sqrt{1 - (b/a)^2} \ .
\end{equation}

\noindent
We will refer to the ratio $b/a \leq 1$ of the shorter and longer
semiaxes of the spheroids as the aspect ratio.

The spheroids are centered at the points $x_n=hn$, where $n$ is an
integer and $h$ is the chain period.  The surface-to-surface
separation of two neighboring spheroids is $\sigma=h-2b$; we require
that $h>2b$ to avoid geometrical intersection of particles. Note that
a stronger condition on the interparticle separation will be imposed
below. The smaller semiaxis $b$ is assumed to be on the order of
$10{\rm nm}$ while $a$ can be up to a few times larger. We will see
that SPPs propagating in such chains have frequencies $\omega$ such
that the corresponding wavelength in the host medium, $\lambda=2\pi
c_h/\omega$, is considerably larger than both $a$ and $b$.
Accordingly, we adopt the dipole approximation.

This choice, however, requires an additional justification. While the
dipole approximation accuracy for electromagnetically interacting
spheroids has not been studied directly, many results are available
for spheres.  The most basic and frequently considered example is that
of two electromagnetically-interacting spheres in close proximity of
each other~\cite{ruppin_89_1,mazets_00_1}. In this case, the spatially
inhomogeneous fields scattered by the spheres result in the excitation
of vector spherical harmonics of all orders, even if the spheres are
small compared to the wavelength. More specifically, the electric
field inside each sphere can be expanded into the vector spherical
harmonics with nonzero coefficients appearing in arbitrarily high
orders. In the dipole approximation, only the first-order terms
($l=1$, $m=0,\pm 1$) are retained in this expansion. The accuracy of
this approximation was found to be dramatically affected by
polarization. If the electric field polarization is parallel to the
axis connecting the spheres, the dipole approximation starts to
deviate from the exact solution when $\sigma \approx 0.5R$, $R$ being
the sphere radius, for both dielectric~\cite{ruppin_89_1} and
conducting~\cite{mazets_00_1} spheres. However, if the polarization is
perpendicular to the axis, the dipole approximation yields results
(i.e., the total dipole moment of the spheres~\cite{mazets_00_1}) with
a relative error of only 2\% even in the case $\sigma=0$. A careful
study~\cite{markel_04_3} of the transversely-polarized electromagnetic
modes of finite-length linear chains of interacting spheres has
revealed that the effect of multipole interaction is to slightly shift
and broaden the dipole resonance - an effect hardly observable in most
materials due to the spectral line broadening associated with Ohmic
losses.

Below, we work in the regime when $\sigma = 2b$ ($h=4b$). In the case
of spheres ($b=R$), one could expect the dipole approximation to be
very accurate for such relative separations regardless of
polarization. We can, however, apply a more stringent test and compare
$\sigma$ to the larger semiaxis of the spheroid, $a$.  Within the
dipole approximation, the physical effect described below is manifest
for the following aspect ratios. For transversely polarized SPP, the
gap in the first Brillouin zone of the chain lattice appears when $b/a
\lesssim 0.25$ in the case of prolate spheroids and when $b/a \lesssim
0.35$ in the case of oblate spheroids. For longitudinal SPP
polarization, the gap appears when $b/a \lesssim 0.1$ in the case of
prolate spheroids and when $b/a \lesssim 0.25$ in the case of oblate
spheroids.

Consider first the transverse polarization. The aspect ratio
$b/a=0.25$ corresponds to $\sigma/a = 0.5$. In the case of two spheres
of radius $R$ separated by the surface-to-surface distance
$\sigma=0.5R$, the multipole effects are negligible. In fact, even a
smaller aspect ratio of $b/a=0.15$, which is used in the numerical
simulations of transversely-polarized wave packets in
Section~\ref{sec:packets}, corresponds to the relative separation
$\sigma=0.3R$. At this separation, the multipole effects can still be
safely ignored. In the case of the longitudinal polarization, the
aspect ratio which is required to observe the effect in chains of
prolate spheroids is $0.1$ which corresponds to $\sigma = 0.2R$. The
multipole effects in this situation are expected to be significant but
not dramatic. However, in oblate spheroid chains, the required aspect
ratio is $0.25$ which corresponds to $\sigma = 0.5R$. At this relative
separation, the effects of higher multipoles are noticeable but small,
even for the longitudinal polarization.

Finally, a simple physical explanation for the dramatic polarization
dependence of the dipole approximation accuracy is available.  When
the polarization is parallel to a chain of spheres, the sphere
surfaces which are adjacent to the "junctions" are similar to usual
capacitors and acquire large and sign-opposite surface charge
densities which, in turn, results in highly non-uniform, strongly
enhanced local fields. This causes excitation of very high multipole
moments. However, for the case of transverse polarization, the surface
charge densities near the junctions are proportional to the
geometrical factor $\cos\theta$ ($\theta$ being the angle between the
polarization vector and the radius vector of a point on the sphere
surface) and are small. Obviously, this consideration holds for
spheroids as well.

We thus conclude that the use of the dipole approximation is well
justified for the purpose of this paper. In the case of transverse SPP
polarization, the approximation accuracy is exceedingly good. For the
longitudinal polarization, the accuracy can be questioned in prolate
spheroid chains but is quite reasonable when oblate spheroids are
used. We will report numerical computation of the dispersion curves
for both prolate and oblate spheroids, in transverse and longitudinal
polarization, and for various aspect ratios of the spheroids
(Section~\ref{sec:dispersion},
Figs.~\ref{fig:disp_prol_ort},\ref{fig:disp_prol_par},\ref{fig:disp_obl_ort},\ref{fig:disp_obl_par}).
However, direct simulation of wave packet propagation
(Sections,\ref{sec:packets},
Figs.~\ref{fig:packet_ort},\ref{fig:packet_par}) is reported only for
the choice of parameters such that the accuracy of the dipole
approximation is not in doubt.

In the dipole approximation, each nanoparticle is characterized by a
(possibly, tensor) dipole polarizability $\alpha(\omega)$ and radiates
as a point dipole. The Cartesian components of the nanoparticle dipole
moments $d_n$ are coupled to each other and to the external electric
field by the coupled-dipole
equation~\cite{markel_07_2,markel_93_1,markel_95_1,markel_05_2}, which
we write here in the frequency domain as

\begin{equation}
\label{CDE}
d_n = \alpha(\omega)\left[E_n^{\rm ext} + \sum_{m\neq n} G_k(x_n,x_m) d_m \right] \ .
\end{equation}

\noindent
Here $E_n^{\rm ext}$ is the external field amplitude at the $n$-th
site, $k=\omega/c_h$ is the wave vector in the host material at the
frequency $\omega$ and $G_k(x,x^{\prime})$ is the appropriate element
of the frequency-domain free-space Green's tensor.  In this paper, we
consider both the transverse and the longitudinal polarizations of the
SPP. In the absence of magnetic polarizability of the nanoparticles
(which is assumed), the SPPs with the three orthogonal polarizations
are not electromagnetically coupled to each other.  Therefore, each
polarization can be considered separately and the quantities appearing
in Eq.~\eqref{CDE} should be understood as follows: $d_n$ and
$E_n^{\rm ext}$ are projections of the dipole moments and of the
external electric field on the selected polarization axis,
$\alpha(\omega)$ is the appropriate scalar element of the
polarizability tensor and $G_k(x,x^{\prime})$ is defined by

\begin{equation}
\label{G_def_a}
G_k(x,x^{\prime}) = \left\{ 
\begin{array}{l}
\left(\frac{\displaystyle k^2}{\displaystyle \vert x -
    x^{\prime}\vert} + \frac{\displaystyle ik}{\displaystyle \vert x -
    x^{\prime}\vert^2} - \frac{\displaystyle 1}{\displaystyle \vert x
    - x^{\prime}\vert^3}\right) \exp(ik\vert x - x^{\prime}\vert) \ ,
\ \ \text{transverse polarization} \ , \vspace*{3mm} \\ 
  2\left(-\frac{\displaystyle ik}{\displaystyle \vert x -
      x^{\prime}\vert^2} + \frac{\displaystyle 1}{\displaystyle \vert x -
      x^{\prime}\vert^3}\right) \exp(ik\vert x -x^{\prime}\vert) \ , \
  \ \text{longitudinal polarization} \ . 
\end{array} \right.
\end{equation}

\section{The Dispersion Relations}
\label{sec:dispersion}

An SPP mode is an excitation that propagates along the chain without
an external source. Thus, to find the dispersion relation, we seek a
solution to \eqref{CDE} with zero free term, $E_n^{\rm ext}=0$, in the
form $d_n \propto \exp(iq x_n)$, where $q$ is in the first Brillouin
zone of the lattice, $q\in [-\pi/h, \pi/h]$.  Substitution of this
ansatz into \eqref{CDE} yields the equation

\begin{equation}
\label{Disp_gen}
\alpha^{-1}(\omega) = h^{-3} S(hk, hq) \ ,
\end{equation}

\noindent
where $S(hk,hq)$ is the dimensionless dipole sum (the dipole
self-energy) of the chain defined by

\begin{equation}
\label{Dip_Sum}
S(\xi, \eta) = \left\{ \begin{array}{l}
2\xi^3{\displaystyle \sum_{n=1}^{\infty}} \left[\frac{\displaystyle 1}{\displaystyle
    n\xi} + \frac{\displaystyle i}{\displaystyle (n\xi)^2} -
  \frac{\displaystyle 1}{\displaystyle (n\xi)^3}\right]
  \exp(i n \xi) \cos( n \eta) \ , \ \ \text{transverse polarization} \
  , \vspace*{3mm} \\
4\xi^3{\displaystyle \sum_{n=1}^{\infty}} \left[-\frac{\displaystyle i}{\displaystyle
     (n\xi)^2} + \frac{\displaystyle 1}{\displaystyle (n\xi)^3}\right]
  \exp(i n \xi) \cos( n \eta) \ , \ \ \text{longitudinal polarization}
  \ .
\end{array} \right.
\end{equation}

\noindent
Note that the dipole sum is a function of two dimensionless parameters
$\xi=kh$ and $\eta=qh$ but does not depend on the particle shape and
material properties.

The dispersion relation, i.e., the mathematical dependence of the SPP
frequency $\omega = k c_h$ on its Bloch wave number $q$, can be
obtained by finding all pairs of variables $(\omega,q)$ that satisfy
Eq.~\eqref{Disp_gen}. This can give rise to one or more branches of
the complex function $\omega(q)$. Generally, purely real pairs
$(\omega,q)$ that solve \eqref{Disp_gen} do not exist. One can,
however, consider purely real values of $q$ and seek complex
frequencies, as was done by Koenderink and
Polman~\cite{koenderink_06_1}. The imaginary part of $\omega$ is then
interpreted as the SPP decay rate. Alternative approaches include
numerical computation of discrete modes in a finite
chain~\cite{weber_04_1} and plotting ${\rm Im}[\alpha^{-1} - h^{-3}
S]^{-1}$ as a function of two variables $k$ and $q$ and visually
identifying the points at which this function appears to have a
maximum or a saddle point~\cite{fung_07_1}.

In this paper, we are interested in propagation of wave packets which
are excited as superpositions of oscillations with purely real
frequencies. Since SPPs propagate without radiative losses, their
Bloch wave numbers $q$ are purely real if the chain is made of a
non-absorbing material such as an ideal conductor. However, when Ohmic
losses in realistic metal are accounted for, $q$ acquires an imaginary
part. In what follows, we use two different approaches to computing
the dispersion curve. In Section~\ref{subsec:real_q}, we consider
chains made of ideal (lossless) metal and seek purely real solutions
$\omega(q)$, as was suggested by Simovski~\cite{simovski_05_1}.
Numerically, this is accomplished by finding pairs of real variables
$(\omega,q)$ that satisfy the dispersion equation~\eqref{Disp_gen} by
the method of bisection. Such purely real solutions exist if the
permittivity of metal is taken to be real.  Next, in
Section~\ref{subsec:real_q}, we consider realistic metals.  Here we
seek pairs $(\omega,q)$ that satisfy the dispersion equation such that
$\omega$ is purely real but $q$ is complex.  Numerically, such pairs
are obtained by utilizing the root-finding algorithm implemented in
Wolfram's {\em Mathematica}. For the specific case of silver, we find
that both approaches yield the results which are very close
quantitatively when the dependence of $\omega$ on ${\rm Re}(q)$ is
considered; the first approach, however, provides no information on
SPP attenuation which is governed by ${\rm Im}q$. In both cases, we
seek solutions only in the region ${\rm Re}q > k = \omega/c_h$; as was
discussed in the Introduction, excitations with ${\rm Re}q < k$
experience radiative decay in addition to Ohmic losses and are not
considered in this paper.

\subsection{Dispersion Relations for Ideal Metal}
\label{subsec:real_q}

In this Section we assume that $q$ and $k$ are real and view the
dipole sum $S(kh,qh)$ as a function of two purely real variables.  As
the first step, we write the inverse polarizability in
\eqref{Disp_gen} in the form~\cite{markel_92_1}:

\begin{equation}
\label{alpha_total}
\alpha^{-1}(\omega)=\alpha_{\rm LL}^{-1}(\omega)-2ik^3/3 \ ,
\end{equation}

\noindent
where $\alpha_{\rm LL}(\omega)$ is the Lorentz-Lorentz quasistatic
polarizability of nanoparticles and $2ik^3/3=i(2/3)(\omega/c_h)^3$ is
the first non-vanishing radiative correction to the inverse
polarizability. The latter is given by

\begin{equation}
\label{z_spheroid}
\alpha^{-1}_{\rm LL}(\omega) = \frac{4\pi}{\epsilon_h v} \left(\nu +
  \frac{\epsilon_h}{\epsilon_m - \epsilon_h} \right) \ ,
\end{equation}

\noindent
where $\nu$ is the appropriate depolarization factor and $v$ is the
spheroid volume~\cite{bohren_book_83}. 

In the case of prolate spheroids, the volume is given by

\begin{equation}
\label{v_prol}
v = \frac{4\pi}{3} ab^2 
\end{equation}

\noindent
and the three depolarization factors are $\nu_1$ for polarization
along the spheroid axis of symmetry and $\nu_2=\nu_3 = (1-\nu_1)/2$
for two linearly independent transverse polarizations, where

\begin{equation}
\label{nu_def_prol}
\nu_1 = \frac{1 - e^2}{e^2} \left[ -1 + \frac{1}{2e}\ln\frac{1 + e}{1 -
    e} \right] \ .
\end{equation}

\noindent
Here $\epsilon_m$ and $\epsilon_h$ are the permittivities of the
(metallic) spheroids and of the host medium, respectively, and $e$ is
the spheroid eccentricity given by formula (\ref{ecc_def}). 

For oblate spheroids, the volume is

\begin{equation}
\label{v_obl}
v = \frac{4 \pi}{3} a^2b
\end{equation}

\noindent
and the depolarization factors are $\nu_1=\nu_2$ for two linearly
independent polarizations which are orthogonal to the spheroid axis of
symmetry and $\nu_3 = 1 - 2\nu_1$ for the polarization along the axis
of symmetry, where

\begin{equation}
\label{nu_def_obl}
\nu_1 = \frac{g(e)}{2e^2}\left[\frac{\pi}{2} - \arctan g(e) \right] -
\frac{g^2(e)}{2} \ , \ \ g(e) = \frac{\sqrt{1-e^2}}{e} \ .
\end{equation}

While both permittivities $\epsilon_m$ and $\epsilon_h$ have, in
general, some frequency dependence, here we neglect the dispersion in
the host and assume that $\epsilon_h={\rm const}>0$. Then, for
nanoparticles made of a lossless material, ${\rm Im}(\alpha^{-1}_{\rm
  LL}) = 0$.  At the same time, if $q>k$, the imaginary part of the
dipole sum is~\cite{markel_07_2,burin_04_1} ${\rm Im}[S(kh,qh)] =
-2(kh)^3/3$ and, in the region of $(k,q)$ which is of interest to us,
${\rm Im}\left[\alpha^{-1} - h^{-3}S\right] = 0$. Therefore, the
imaginary part of Eq.~\eqref{Disp_gen} is satisfied identically and
only its real part needs to be considered. We then utilize
Eqs.~\eqref{alpha_total},\eqref{z_spheroid} and arrive at the
following dispersion equation:

\begin{equation}
\label{Disp_Real_alpha}
\nu + {\rm Re}\left(\frac{\epsilon_h}{\epsilon_m - \epsilon_h} \right) = \epsilon_h
\frac{v}{4\pi h^3} {\rm Re}[S(hk, hq)] \ .
\end{equation}

\noindent
In the case of a lossless metal and a transparent host medium, the
real-part symbol in the left-hand side of \eqref{Disp_Real_alpha} can
be omitted.

\begin{figure}
  \centerline{\epsfig{file=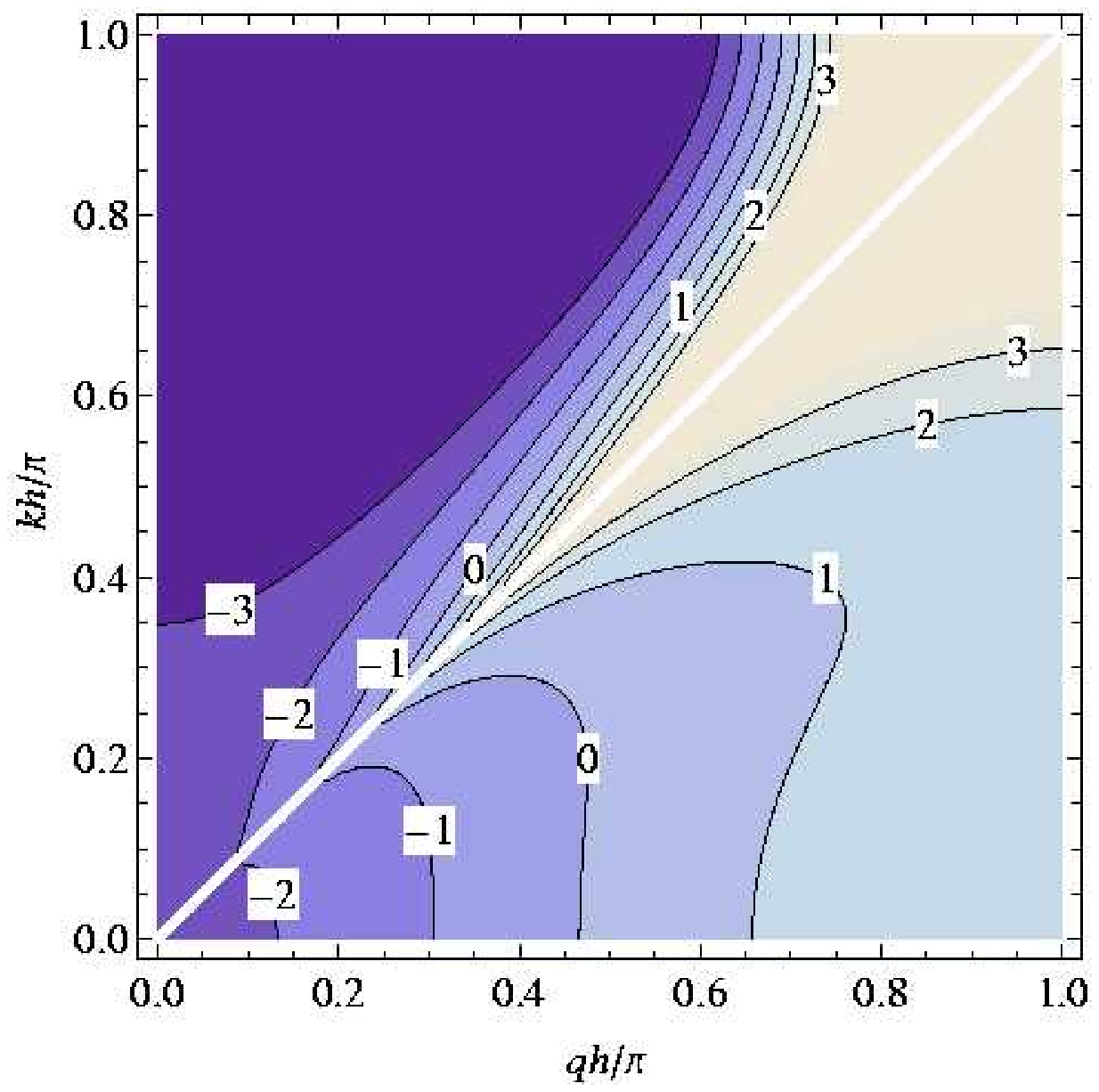,width=8cm,clip=t}}
  \vspace*{3mm}
  \centerline{\epsfig{file=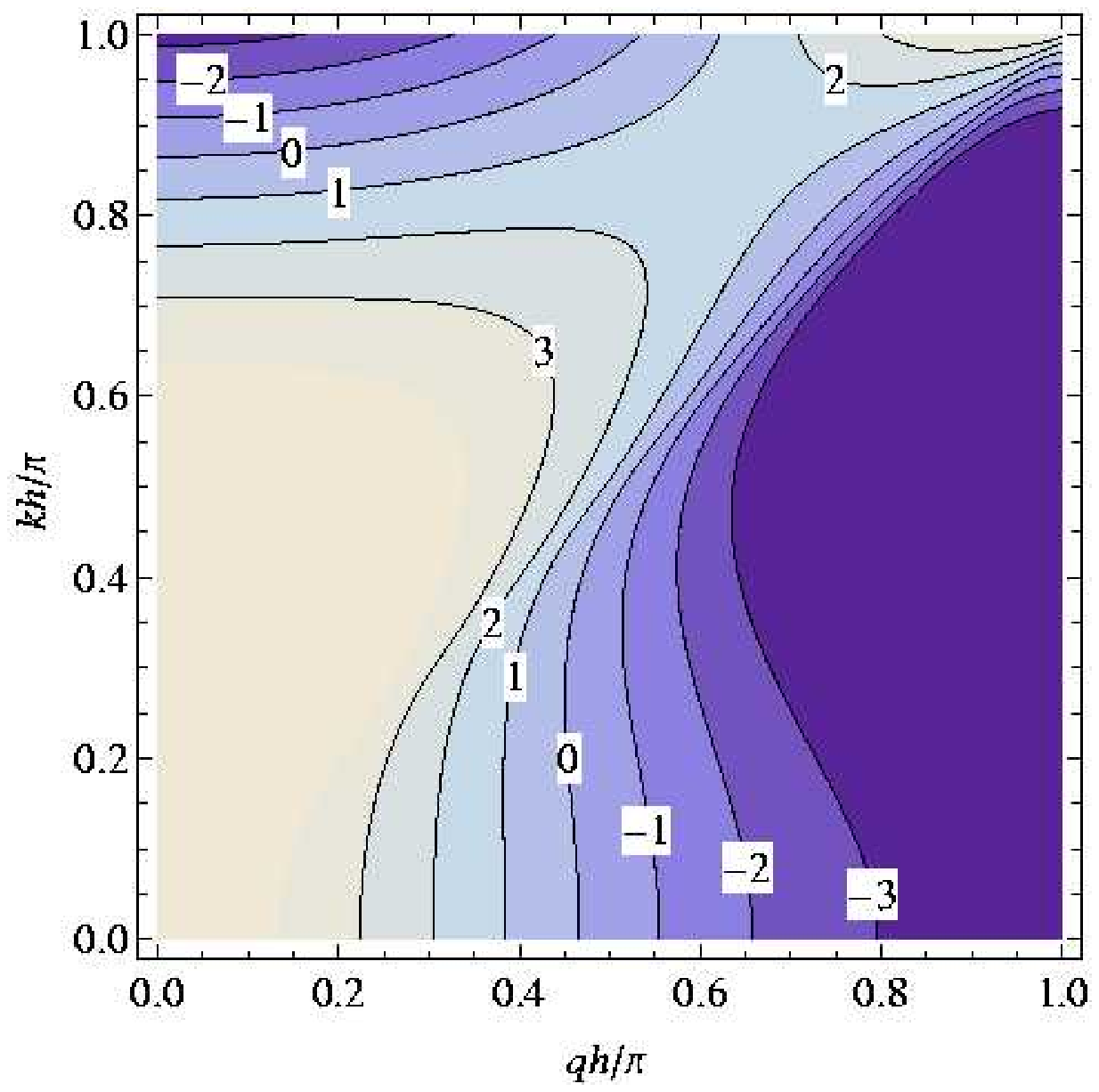,width=8cm,clip=t}}
\caption{\label{fig:dipsum}
  (color online) Contour plot of the real part of the dipole sum,
  ${\rm Re}[S(kh, qh)]$ for the transverse (top) and the longitudinal
  (bottom) polarizations. Due to the symmetry
  $S(\xi,\eta)=S(\xi,-\eta)$, only the positive half of the first
  Brillouin zone is shown.  Note that, for the transverse
  polarization, ${\rm Re}[S(kh, qh)]$ diverges logarithmically
  (approaches positive infinity) on the light line $q=k$. Near this
  line, the function changes so fast that it is not feasible to depict
  it quantitatively using the contour plot; the region of divergence
  is schematically shown as a diagonal white line in the top panel.
  There is no divergence for the longitudinal polarization.}
\end{figure}

Eq.~\eqref{Disp_Real_alpha} allows one to analyze the relation between
the spheroid aspect ratio and the dispersive properties of the LPCs.
Consider first the SPP polarization which is transverse to the chain.
For this polarization, the dependence of the real part of the dipole
sum $S(kh,qh)$ on its arguments is illustrated in
Fig.~\ref{fig:dipsum}(a) and the relevant depolarization factor is
$\nu_1$ (for both prolate and oblate spheroids).  We now notice the
following. In the spectral region where the metal experiences
anomalous dispersion, the second term in the left-hand side of
\eqref{Disp_Real_alpha} is negative, assuming that the host medium is
a transparent dielectric.  When the aspect ratio $b/a$ is decreased,
the depolarization factors $\nu_1$ given by Eqs.~(\ref{nu_def_prol})
or (\ref{nu_def_obl}), for prolate and oblate spheroids, respectively,
both approach zero.  As a result, the whole left-hand side in
\eqref{Disp_Real_alpha} becomes negative.  On the other hand, there
are regions in the $(k,q)$ space in which the right-hand side of
\eqref{Disp_Real_alpha} is strictly positive.  Thus, it can be seen
from Fig.~\ref{fig:dipsum}(a) that ${\rm Re}[S(kh, qh)]>0$ if $qh/\pi
\gtrsim 0.5$ and $q>k$.  As a result, for sufficiently small ratio
$b/a$, equation \eqref{Disp_Real_alpha} ceases to have real-valued
solutions if $q > q_c^\perp$, where $q_c^\perp$ is the (aspect
ratio-dependent) critical value of $q$ specific to the transverse
polarization. The interval of Bloch wave numbers $q > q_c^\perp$
corresponds to a gap in the first Brillouin zone of the chain lattice
in which SPPs do not exist. It is important to emphasize that the
critical constant $q_c^\perp < \pi/h$ exists only for sufficiently
small ratio $b/a$. 

Similar considerations can be applied to the longitudinal SPP
polarization, except that the relevant depolarization coefficient is,
in this case, $\nu_3$. The dependence of ${\rm Re}[S(kh,qh)]$ on its
arguments is illustrated in Fig.~\ref{fig:dipsum}(b). It can be seen
that ${\rm Re}[S(kh,qh)]$ is strictly positive for $qh/\pi \lesssim
0.4$ and $q>k$. Correspondingly, Eq.~\eqref{Disp_Real_alpha} ceases to
have real-valued solutions for $q < q_c^\parallel$ which
defines a gap in the first Brillouin zone of the lattice.  Here
$q_c^\parallel$ is the critical Bloch wave number for the parallel
polarization; as in the case of transverse polarization,
$q_c^\parallel$ exists only for sufficiently small aspect ratios and
is aspect ratio-dependent. Note that the depolarization factor $\nu_3$
approaches a finite value rather than zero when the aspect ratio is
decreased. This limit is $1/2$ for prolate and $1$ for oblate
spheroids. Due to this reason, observation of the gap for
longitudinally-polarized SPPs requires a smaller aspect ratio. This
point will be illustrated below by numerical examples.

We now compute the dispersion curves numerically.  To solve
Eq.~\eqref{Disp_Real_alpha}, a specific expression for the metal
permittivity $\epsilon_m$ is needed. We use here the Drude formula

\begin{equation}
\label{drude}
\epsilon_m = \epsilon_0 - \frac{\omega_p^2}{\omega(\omega + i\gamma)}
\ ,
\end{equation}

\noindent
where $\epsilon_0$ is the contribution due to interzone
transitions~\cite{kreibig_85_1}, $\omega_p$ is the plasma frequency,
and $\gamma$ is the relaxation constant. In this subsection, we set
$\gamma=0$ to describe a lossless metal (realistic values of $\gamma$
will be used in Sections~\ref{subsec:complex_q} and \ref{sec:packets}
below). We then solve Eq.~\eqref{Disp_Real_alpha} by the method of
bisection to obtain the dispersion curves and the SPP group velocity.

\begin{figure}
\centerline{\psfig{file=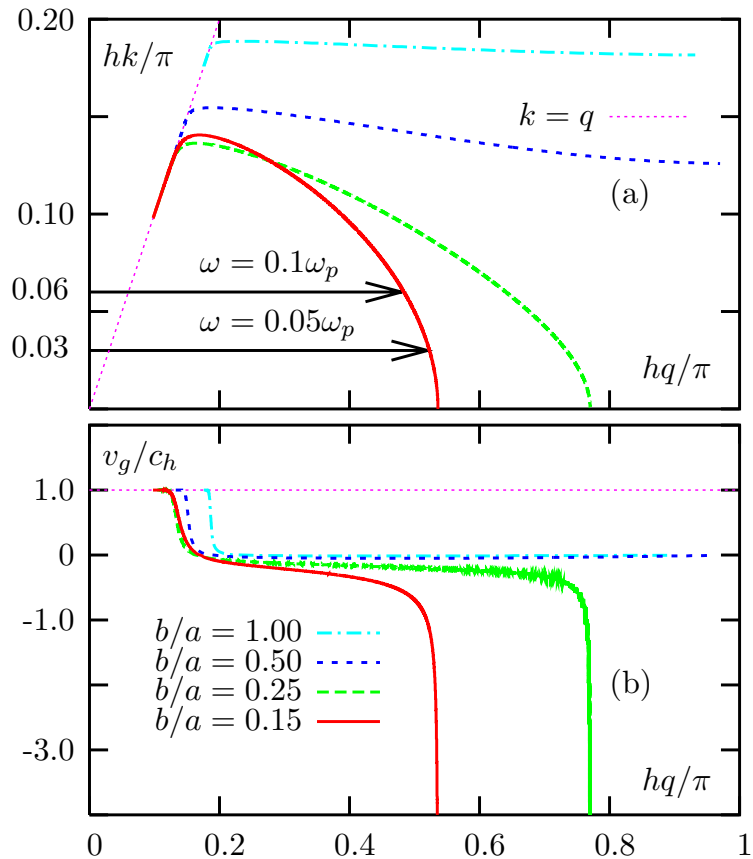,width=11cm,bbllx=170,bblly=475,bburx=450,bbury=770,clip=}}
\caption{\label{fig:disp_prol_ort} (color online) Dispersion curves (a) and group
  velocity (b) for transversely polarized SPPs in chains built from
  prolate spheroids whose axis of symmetry is perpendicular to the
  chain, for different spheroid aspect ratios $b/a$ and for fixed
  ratios $h/b=4$ and $\lambda_p/h=3.4$. Since the dispersion curves
  are symmetric with respect to the $k$-axis, only the positive half
  of the first Brillouin zone ($q>0$) is shown. Only data points that
  were found numerically are plotted. Theoretically, however, all
  dispersion curves in infinite chains start at the point $k=q=0$ and
  follow the light line (labeled as $k=q$ in the top panel) for some
  range of $q$'s. Horizontal arrows indicate central frequencies of
  the Gaussian wave packets whose simulated propagation is illustrated
  in Fig.~\ref{fig:packet_ort} below.}
\end{figure}

\begin{figure}
\centerline{\psfig{file=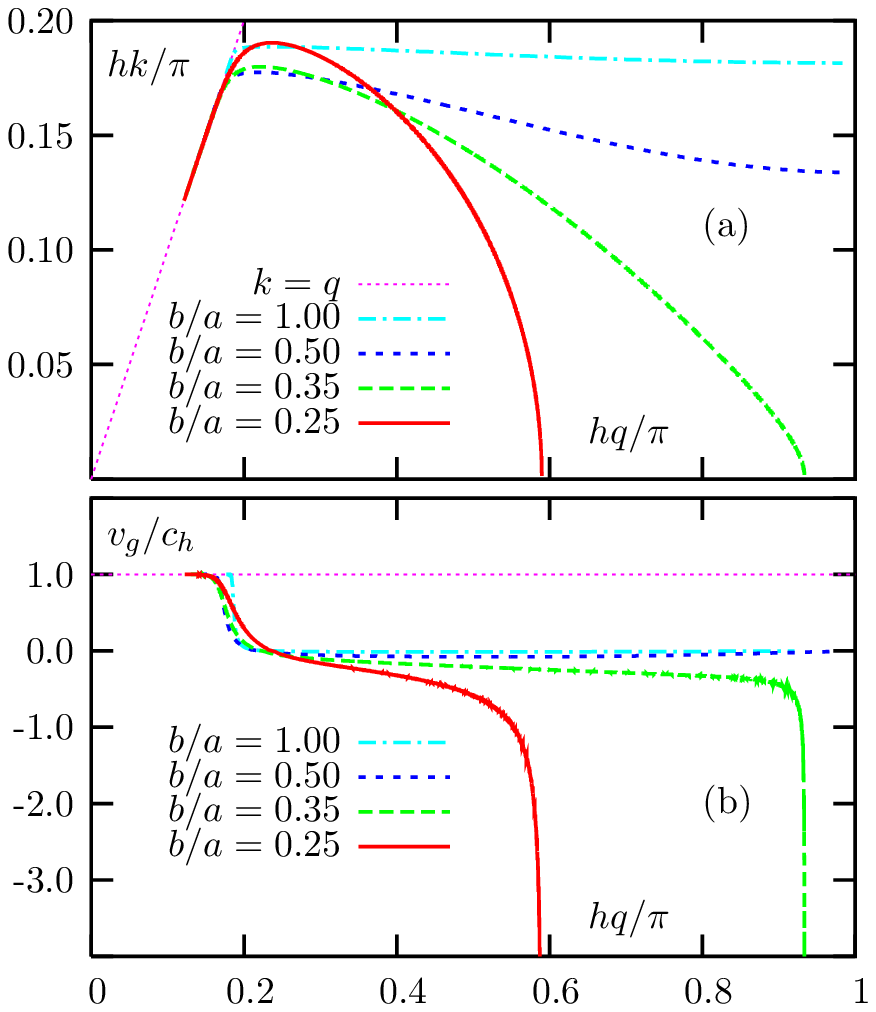,width=11cm,bbllx=170,bblly=475,bburx=450,bbury=770,clip=}}
\caption{\label{fig:disp_obl_ort} (color online) Same as in
  Fig.~\ref{fig:disp_prol_ort} but for a chain made of oblate
  spheroids whose axis of symmetry is parallel to the chain and for a
  different set of aspect ratios. SPP polarization is orthogonal to
  the chain.}
\end{figure}

\begin{figure}
  \centerline{\psfig{file=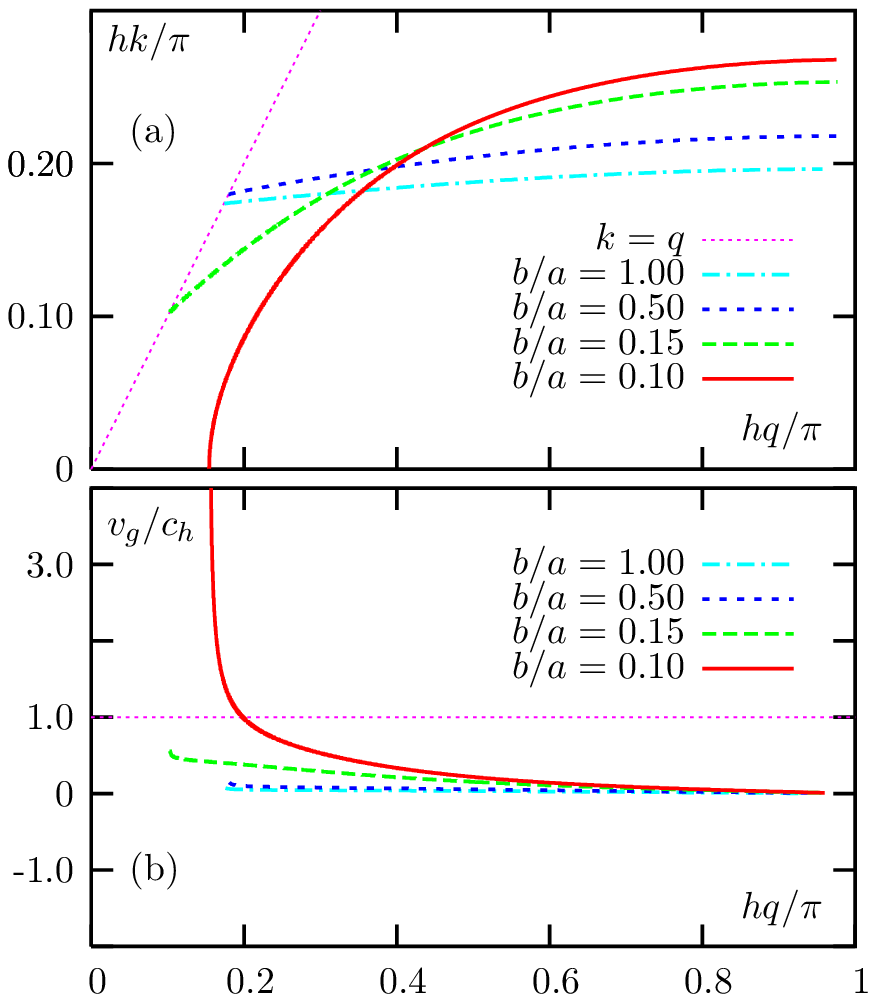,width=11cm,bbllx=170,bblly=475,bburx=450,bbury=770,clip=}}
\caption{\label{fig:disp_prol_par} Same as in
  Fig.~\ref{fig:disp_prol_ort} but for longitudinal SPP polarization
  and a different set of aspect ratios. Note that, unlike in the case
  of transverse polarization, the dispersion curves do not have linear
  small-$q$ segments. As stated in the text, only data points that
  satisfy $q\geq k$ are plotted, since there are no real-valued
  solutions in the region $q<k$.}
\end{figure}

\begin{figure}
\centerline{\psfig{file=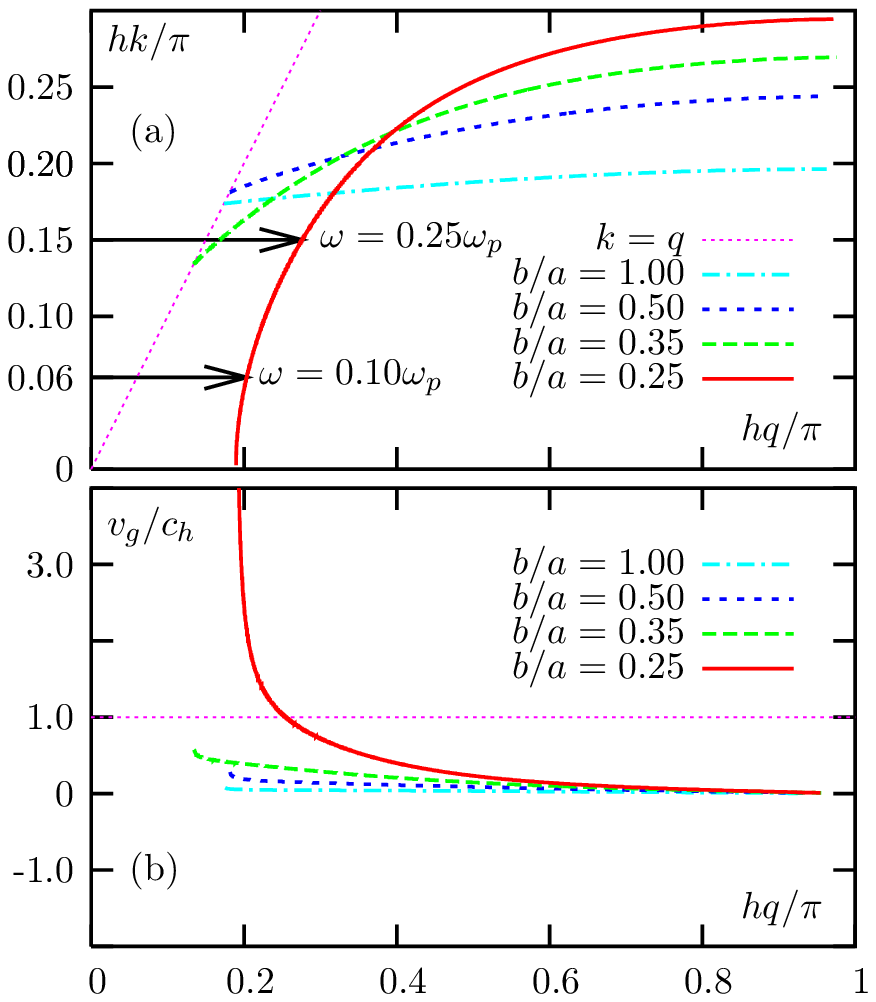,width=11cm,bbllx=170,bblly=475,bburx=450,bbury=770,clip=}}
\caption{\label{fig:disp_obl_par} Same as in
  Fig.~\ref{fig:disp_obl_ort} but for longitudinal SPP polarization
  and a different set of aspect ratios. Note that, unlike in the case
  of transverse polarization, the dispersion curves do not have linear
  small-$q$ segments. As stated in the text, only data points that
  satisfy $q\geq k$ are plotted, since there are no real-valued
  solutions in the region $q<k$.  Horizontal arrows indicate central
  frequencies of the Gaussian wave packets whose simulated propagation
  is illustrated in Fig.~\ref{fig:packet_par} below.}
\end{figure}

The results are shown in Figs.~\ref{fig:disp_prol_ort} through
\ref{fig:disp_obl_par}. The top panels in these four figures show a
number of dispersion curves computed for different aspect ratios
$b/a$, as labeled, and for different SPP polarizations. The group
velocity of the SPPs, $v_g = \partial \omega/ \partial q$, is plotted
as a function of $q$ in the bottom panels. The values of $\epsilon_h$
and $\epsilon_0$ are taken to be $2.5$ (as in the case of a glassy
medium) and $5.0$ (the experimental value for
silver~\cite{kreibig_85_1}), respectively. The computation does not
depend on the absolute value of the plasma frequency $\omega_p$ but on
the dimensionless parameter $\lambda_p/h$, where $\lambda_p = 2\pi
c_h/\omega_p$ is the wavelength (in the host medium) at the plasma
frequency. This ratio was chosen to be $\lambda_p/h = 3.4$.  Finally,
we have set the ratio $h/b=4$ so that the minimum surface-to-surface
separation of two neighboring spheroids, $\sigma$, was equal to $2b$.
As an illustration, for the specific case of silver, we have the
following parameters: the vacuum plasma wavelength is $\lambda_p^{\rm
  (vac)} \approx 136 {\rm nm}$ and the corresponding value in the host
medium is $\lambda_p = \lambda_p^{\rm (vac)}/\sqrt{\epsilon_h}\approx
86{\rm nm}$; correspondingly, $h\approx 25{\rm nm}$, $b \approx 6{\rm
  nm}$ and $a$ varies from $6 {\rm nm}$ to $60{\rm nm}$. The latter
value was obtained for the smallest aspect ratio used, $b/a=0.1$.  It
can be seen that, for all points on the dispersion curves shown in
Figs.~\ref{fig:disp_prol_ort}-\ref{fig:disp_obl_par}, SPP frequencies
are well below the plasma frequency $\omega_p$ and $kh/\pi \ll 1$. The
last inequality is especially strong for the central frequencies
(indicated by horizontal arrows in Figs.~\ref{fig:disp_prol_ort} and
\ref{fig:disp_obl_par}) which were used to simulate wave packet
propagation in Section~\ref{sec:packets} below. This reconfirms the
dipole approximation validity.

We now discuss the computed dispersion curves in more detail, starting
with the case of transverse SPP polarization
(Figs.~\ref{fig:disp_prol_ort},\ref{fig:disp_obl_ort}). First, we note
that in infinite, strictly periodic chains, the dispersion curves of
transversely-polarized SPPs start at the point $k=q=0$ and then follow
the light line $k=q$ for some range of $q$. This small-$q$ part of the
dispersion curve is related to the logarithmic divergence of the
dipole sum on the light line~\cite{markel_05_2} and is extremely
difficult to find numerically. Indeed, in order to satisfy
Eq.~\eqref{Disp_Real_alpha}, the points on the small-$q$ section of
the dispersion curve must be specified with exponentially large
numerical precision. This is why the small-$q$ section of the
dispersion curve has not been reported in a number of numerical
investigations~\cite{weber_04_1,simovski_05_1,koenderink_06_1,fung_07_1}.
However, the SPPs with $q\approx k \ll \pi/h$ exist and were observed
in numerical simulations~\cite{markel_07_2}.

In this paper, we do not consider the small-$q$ part of the dispersion
curve but focus on the SPPs for which the ratio $k/q$ is considerably
less than unity. Such modes exist for $qh/\pi \gtrsim 0.2$ for all
four values of $b/a$ shown in
Figs.~\ref{fig:disp_prol_ort},\ref{fig:disp_obl_ort}. The main point
of this paper is that the dispersion curve shape is strongly
influenced by the ratio $b/a$. When $b/a=1.0$, the corresponding
dispersion curve is almost flat (apart from the linear small-$q$
section of the curve). At $b/a=0.5$, the curve begins to bend down
noticeably at larger values of $q$.  Finally, when $b/a \lesssim 0.25$
(in the case of prolate spheroids) or when $b/a \lesssim 0.35$ (in the
case of oblate spheroids), the curves cross the $k=0$ axis at the
point $q=q_c^\perp$ and no solutions exist for $q > q_c^\perp$. Near
the critical point, the dispersion curves acquires a very large
negative slope. Note that the corresponding slope is positive in the
$q<0$ part of the first Brillouin zone (which is not shown in the
figures).

From comparison of
Figs.~\ref{fig:disp_prol_ort},\ref{fig:disp_obl_ort}, a conclusion can
be made that the dispersion curves are more sensitive to the aspect
ratio in the case of oblate spheroids. In particular, the gap in the
first Brillouin zone of the chain lattice appears for more moderate
values of $b/a$ if the chain is made of oblate spheroids.

Dispersion curves for the longitudinal SPP polarization are shown in
Figs.~\ref{fig:disp_prol_par},\ref{fig:disp_obl_par}. It can be seen
that, for sufficiently small aspect ratios, there exists a critical
value $q_c^\parallel$ such that Eq.~\eqref{Disp_Real_alpha} has no
real-valued solutions for $q<q_c^\parallel$. Thus, SPP propagation in
the chain is only possible for $q$ larger than the critical value
$q_c^\parallel$ (if the latter exists). The group velocity shown in
Figs.~\ref{fig:disp_prol_par}(b),\ref{fig:disp_obl_par}(b) acquires
large positive values in the vicinity of $q_c^\parallel$.

Similarly to the case of transverse SPP polarization, the dispersion
curves are more sensitive to the aspect ratio if the chains are made
of oblate spheroids. Thus, in the case of prolate spheroids, the gap
in the first Brillouin zone of the lattice appears only when $b/a
\lesssim 0.1$. However, if the chain is composed of oblate spheroids,
the gap appears when $b/a \lesssim 0.25$.

\subsection{Dispersion Relations for Realistic Metal}
\label{subsec:complex_q}

The dispersion curves shown in Figs.~\ref{fig:disp_prol_ort} through
\ref{fig:disp_obl_par} contain no information on either the rate or
the direction of SPP spatial decay. However, it can be seen in these
figures that an SPP is characterized by phase and group velocities and
that these can have different signs when projected onto the $x$-axis.
On physical grounds, we expect that wave packets should decay in the
direction of propagation. This imposes certain restrictions on the
signs of the real and the imaginary parts of $q$. Thus, if $v_g v_p <
0$, we expect that ${\rm Re}(q) {\rm Im}(q) < 0$ while if $v_g v_p >
0$, we expect that ${\rm Re}(q) {\rm Im}(q) > 0$.

The above statement can be illustrated by considering the following
thought experiment. Assume that a short Gaussian optical pulse is
injected into the central part of a long chain by a spatially
localized source such as a near-field microscope tip operating in the
illumination mode. The pulse will propagate in the form of two wave
packets in both directions along the chain. If $v_g v_p < 0$, both
wave packets would be composed of Bloch waves whose phase velocities
point towards the source and group velocities point away from the
source. Since the sign of the phase velocity is the same as that of
${\rm Re}(q)$, and since both wave packets should decay in the
direction of propagation, we expect in this case that ${\rm Re}(q)
{\rm Im}(q) < 0$. Similar consideration can be applied to the case
$v_g v_p > 0$.

In the numerical simulations of this subsection, we have verified that
the above analysis is, indeed, correct. To this end, we have
incorporated the Ohmic losses into the model of metal permittivity.
Specifically, we have set the Drude relaxation constant in
Eq.~\eqref{drude} to be $\gamma = 0.002\omega_p$ which is the
experimental value for silver. Note that Eq.~\eqref{Disp_Real_alpha}
can still be satisfied in this case with a purely real pair of
$(\omega,q)$ (the real-part symbol in the left-hand side of the
equation must be retained if $\epsilon_m$ is complex-valued). However,
the imaginary part of the more general Eq.~\eqref{Disp_gen} is no
longer satisfied identically. We, therefore, must seek such pairs of
variables $(\omega,q)$, where $q$ is now complex, that satisfy both
the real and the imaginary parts of Eq.~\eqref{Disp_gen}. This can not
be achieved by the use of the simple bisection algorithm that was
employed in the previous subsection. Instead, we employ here the
Wolfram's {\em Mathematica} root finder to obtain complex roots of Eq.~\eqref{Disp_gen},
$q$, for each real value of $\omega$.

\begin{figure}
\centerline{\psfig{file=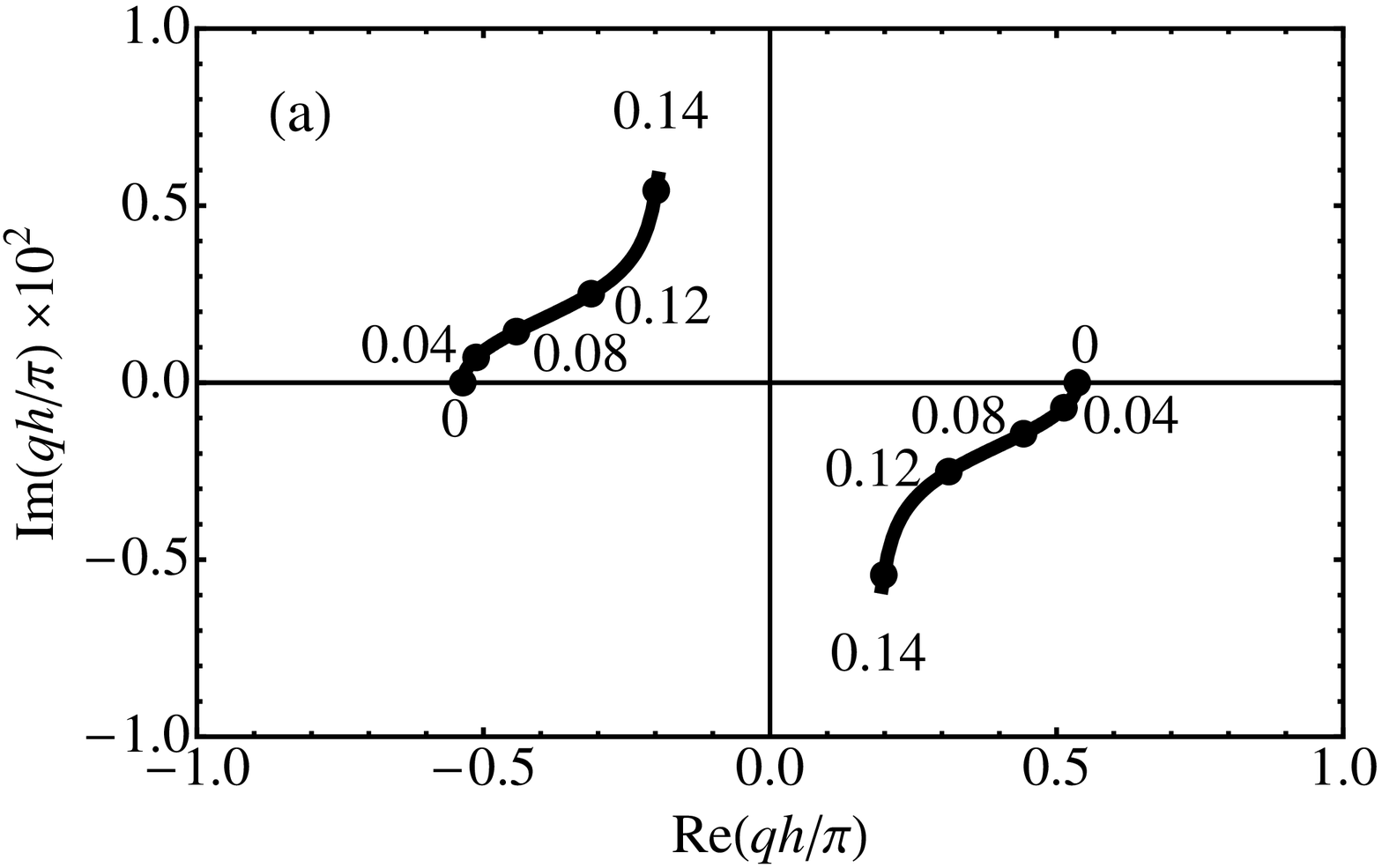,width=11cm,clip=t}}
\vspace*{3mm}
\centerline{\psfig{file=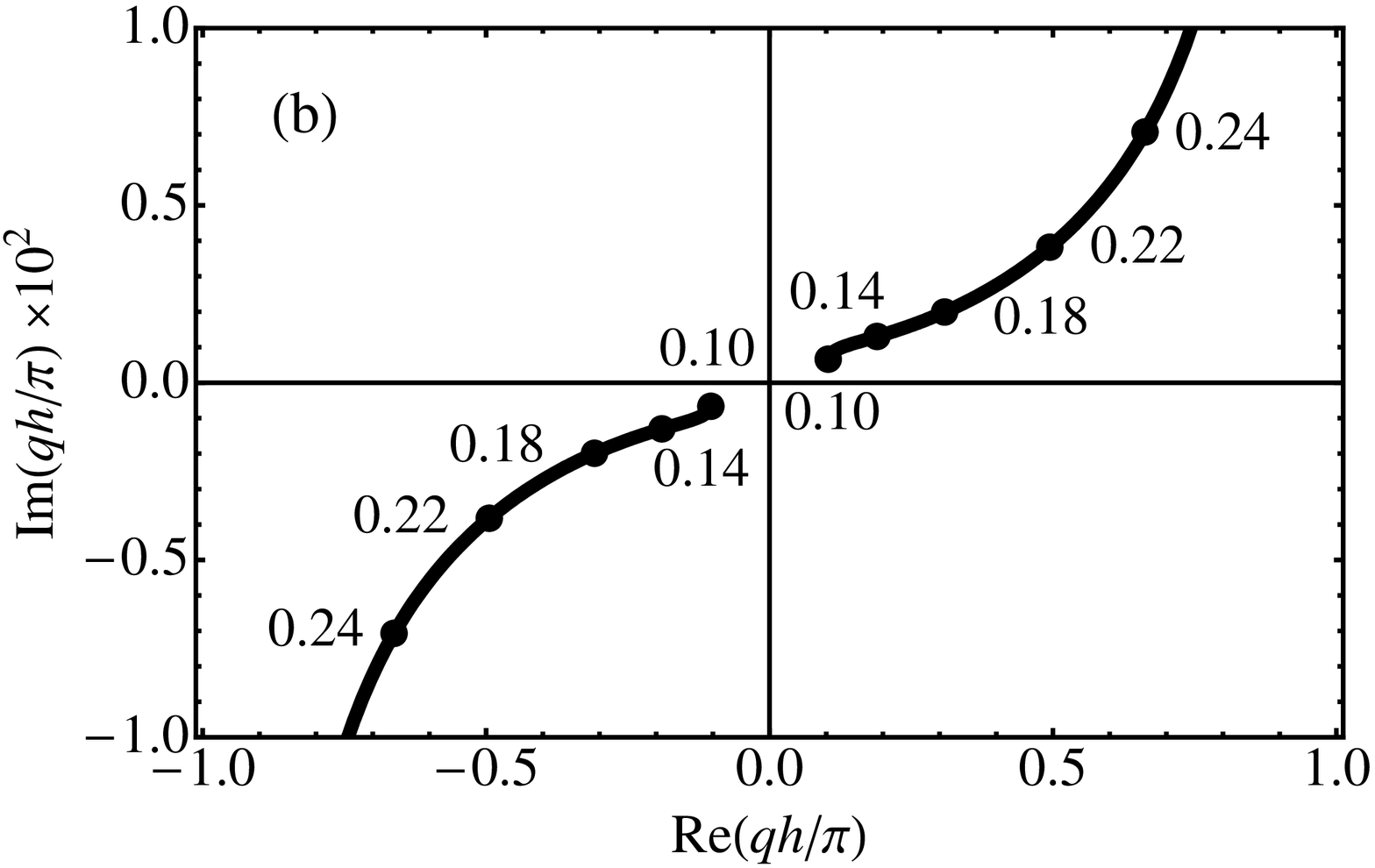,width=11cm,clip=t}}
\caption{\label{fig:disp_prol_ar0.15} Dispersion curves plotted parametrically
  in the complex $q$-plane for transversely (a) and longitudinally (b)
  polarized SPPs in an LPC of prolate spheroids. Parameters:
  $\gamma/\omega_p = 0.002$, $b/a=0.15$ and other parameters same as
  in Fig.~\ref{fig:disp_prol_ort}. Dots label the values of the
  dimensionless parameter $kh/\pi = \omega h /\pi c_h$.  In panel (a),
  only the points that correspond to the the region $v_p v_g < 0$ are
  shown, which corresponds, approximately, to $\vert {\rm Re}(q) \vert
  h/\pi \gtrsim 0.17$.}
\end{figure}

\begin{figure}
\centerline{\psfig{file=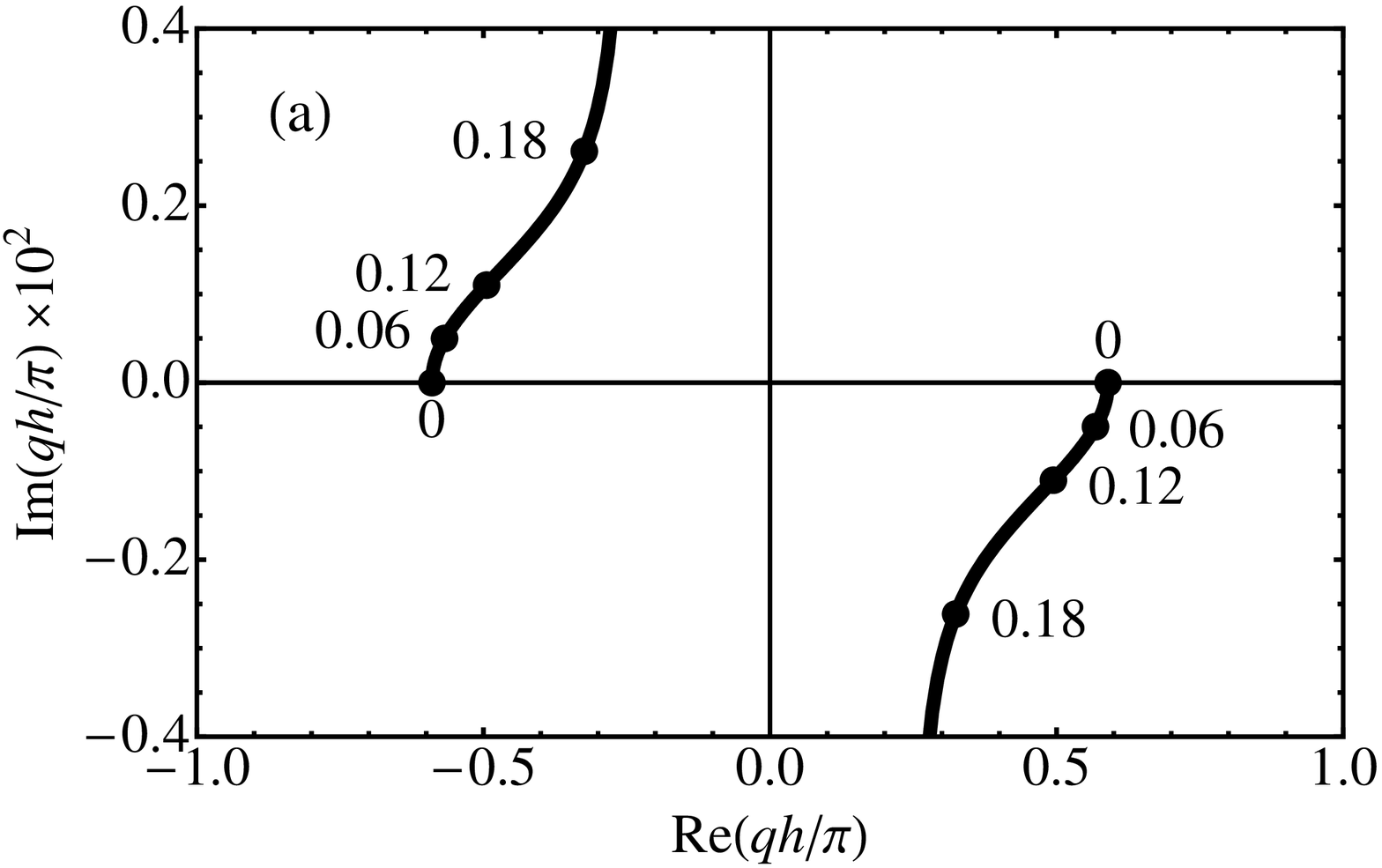,width=11cm,clip=t}}
\vspace*{3mm}
\centerline{\psfig{file=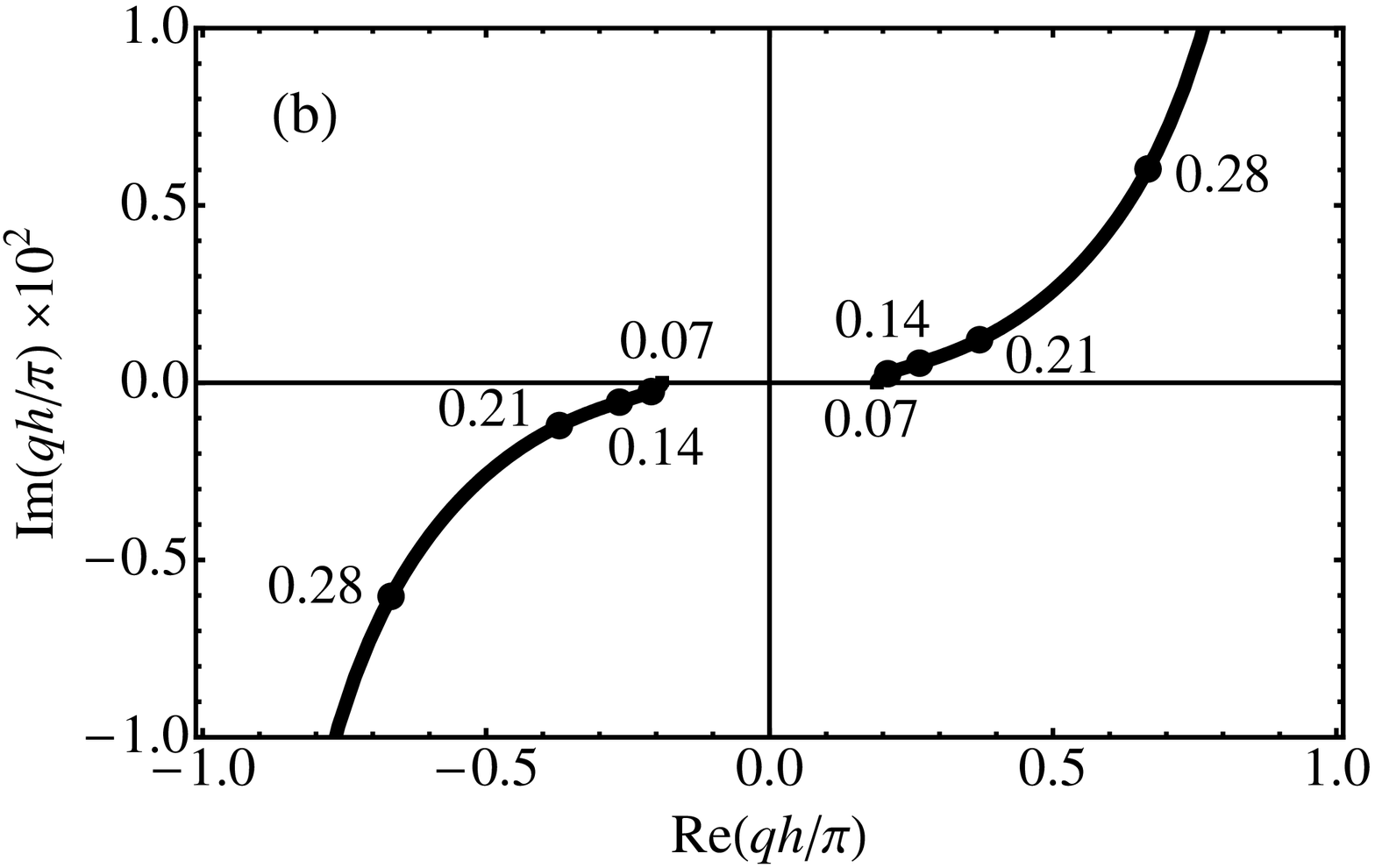,width=11cm,clip=t}}
\caption{\label{fig:disp_obl_ar0.25} Same as in Fig.~\ref{fig:disp_prol_ar0.15}
  for an LPC of oblate spheroids whose aspect ratio is $b/a=0.25$}
\end{figure}

The results of this simulation are plotted parametrically in the
complex $q$-plane in Fig.~\ref{fig:disp_prol_ar0.15} for a chain of
prolate spheroids of the aspect ratio $b/a=0.15$ and in
Fig.~\ref{fig:disp_obl_ar0.25} for a chain of oblate spheroids of the
aspect ratio $b/a=0.25$. Other parameters are the same as in
Fig.~\ref{fig:disp_prol_ort}. As expected, the real and imaginary
parts of the Bloch wave number have opposite signs for the
transversely-polarized SPPs so that the wave packets decay in the
direction of propagation specified by $v_g$, even though the phase
velocity is pointing into the opposite direction. For longitudinally
polarized SPPs the product $v_p v_g$ is always positive and,
correspondingly, ${\rm Im}(q) {\rm Re}(q) > 0$.

An important observation is that when ${\rm Re}q$ approaches one of
its critical values, ${\rm Im}q$ tends to zero. Correspondingly, the
decay length of the SPPs is dramatically increased.

We note that the data shown in Fig.~\ref{fig:disp_prol_ar0.15}
represent essentially the same dispersion curves as the ones plotted
in Fig.~\ref{fig:disp_prol_ort}(a) and \ref{fig:disp_prol_par}(a) for
the case $b/a=0.15$. Similarly, data in
Fig.~\ref{fig:disp_obl_ar0.25} correspond to the dispersion curves
plotted in Fig.~\ref{fig:disp_obl_ort}(a) and
\ref{fig:disp_obl_par}(a) for $b/a=0.25$. The discrepancy between the
curves $\omega[{\rm Re}(q)]$ computed by the two methods is negligibly
small. This further justifies the utility of the simple numerical
method of Section~\ref{subsec:real_q} for computing the dispersion
curves in LPCs.

\section{Attenuation of SPPs and the Decay Length}
\label{sec:l}

The nonzero relaxation constant in Eq.~\eqref{drude} leads to
exponential decay of SPPs even in the absence of radiative losses. The
decay lengths in LPCs is given by $\ell_\textsc{lpc} = 1 / {\rm
  Im}(q)$. If $q$ is not very close to the edge of one of the gaps
that were discussed above, $\ell_\textsc{lpc}$ can be computed in the
quasi-particle pole approximation~\cite{markel_07_2} which results in
the following expression:

\begin{equation}
\label{lChain_gen}
\ell_\textsc{lpc}
 \approx \frac{1}{-h^2{\rm Im}\left[\alpha_{\rm
      LL}^{-1}(\omega)\right]}\left\vert \frac{\partial {\rm
      Re}S(\xi,\eta)}{\partial \eta}\right\vert_{\eta= q_0(\xi)h } 
\end{equation}

\noindent
where $S(\xi,\eta)$ is the dipole sum given by Eq.~\eqref{Dip_Sum} and
$q_0(\xi)$ is a purely real solution to \eqref{Disp_Real_alpha} for a
given value of the dimensionless parameter $\xi = \omega h / c_h$.
Under the assumptions that $\epsilon_h, {\rm Im}(\epsilon_m) \ll {\rm
  Re}(\epsilon_m)$, we have

\begin{equation}
\label{delta_eval}
- {\rm Im} \left[ \alpha_{\rm LL}^{-1} (\omega) \right] \approx 4\pi
\omega\gamma/v \omega_p^2 \ .
\end{equation}

\noindent
As we have seen above, the real-valued solutions to
\eqref{Disp_Real_alpha} satisfy, albeit approximately, the more
general dispersion equation \eqref{Disp_gen}. Therefore, we can
evaluate the derivative in Eq.~\eqref{lChain_gen} at the point
$(\xi,\eta)$ with the understanding that $\xi$ and $\eta$ are purely
real variables that satisfy the approximate dispersion equation
\eqref{Disp_Real_alpha}.  We further neglect the terms that are of the
order of $\sim 1/\xi$ and $\sim 1 /\xi^2$ in the square brackets of
the expression \eqref{Dip_Sum} for the dipole sum and arrive at the
following estimate:

\begin{equation}
\label{lChain}
    \ell_\textsc{lpc} \simeq h\frac{Av}{2\pi h^3}\frac{\omega_p^2}{\omega
      \gamma} \left \vert \sum_{m=1}^\infty
\frac{\cos(\omega h m/c_h )\sin(q h m)}{m^2} \right \vert \ ,
\end{equation}

\noindent
where $A=1$ for the transverse polarization and and $A=2$ for the
longitudinal polarization.

A direct calculation for the transversely-polarized SPP in a
prolate spheroid chain, $b/a=0.15$ and other parameters same as in
Fig.~\ref{fig:disp_prol_ort} yields the decay lengths
$\ell_\textsc{lpc}\simeq 7\mu{\rm m}$ for $\omega = 0.1\omega_p$ and
$\ell_\textsc{lpc}\simeq 15\mu {\rm m}$ for $\omega = 0.05\omega_p$.
Interestingly, the polarization dependence of decay length is
contained solely in the factor $A$.  However, the two polarizations
have markedly different dispersion curves and therefore, same pairs of
variables $(\xi,\eta)$ may not be accessible for the two different
polarizations.

It is instructive to compare the SPP decay length in
nanoparticle chains and metallic nanowires. The dispersion equation in
a metallic cylindrical waveguide is~\cite{govyadinov_06_1,chang_07_1}:

\begin{equation}
\label{disp_wire_exact}
\frac{\epsilon_m I_1(\kappa_m R)}{\kappa_m I_0(\kappa_m
  R)}+\frac{\epsilon_h K_1(\kappa_h R)}{\kappa_h K_0(\kappa_h R)}=0
\end{equation}

\noindent
where $R$ is the cylinder radius, $I_l(x)$ and $K_l(x)$ are the
modified Bessel functions, $\kappa_{m,h}=\sqrt{q^2-\epsilon_{m,h}
  \omega^2/c_h^2}$ and the indices ``$m$'' and ``$h$'' label the
quantities for the metal and for the surrounding dielectric host,
respectively. If the wire is sufficiently thin, we can expand the
Bessel function to the first non-vanishing order in $\kappa_m R$ and
$\kappa_h R$ to obtain the simplified dispersion equation

\begin{equation}
\label{disp_wire_1}
\epsilon_m = \frac{2\epsilon_h}{(\kappa_h R)^2} \frac{1}{{\rm
    ln}(\kappa_h R / 2) + C} \ ,
\end{equation}

\noindent
where $C$ is the Euler constant. This equation can be solved
approximately (with logarithmic precision) as 

\begin{equation}
\label{kappa_sol}
\kappa_h^2 \approx
\frac{2\epsilon_h}{-\epsilon_m}\frac{1}{\displaystyle \frac{1}{2}{\rm
    ln}\frac{-2\epsilon_m}{\epsilon_h} - C} \ .
\end{equation}
 
\noindent
We then use the Drude formula for $\epsilon_m$, take into account the
fact that the propagation constant ${\rm Re}q$ in a metal nanowire is
much larger than $\sqrt{\epsilon_h}\omega/c$ (which is the wavenumber
in the surrounding medium) and obtain the following estimate for the
decay length:

\begin{equation}
\label{lwire}
\ell_{\rm wire} \sim \frac{\omega_p}{\gamma}\frac{R}{2\sqrt{2\epsilon_h}} \ ,
\end{equation}

\noindent
One additional condition that has been used in deriving the above
estimate is $\omega \gg \gamma$. For a silver nanowire of radius $R =
25{\rm nm}$, the estimate yields $\ell_{\rm wire}\approx 2.8\mu {\rm
  m}$.

The following conclusions can be made. While the SPP decay length in
nanowires depends only on the metal and host permittivities and the
waveguide radius, the same quantity in the LPCs can be controlled by
changing the inter-particle separation and the particle dimensions.
Decreasing the inter-particle spacings results in stronger
electromagnetic coupling and larger propagation distances.  However,
for sufficiently small values of $h$, the dipole approximation breaks
down and the estimate~\eqref{lChain} becomes invalid. The dimensions
of a nanoparticle provide another set of degrees of freedom in
controlling the decay length in LPCs. Note that the expression
\eqref{lChain} contains an overall factor $v/ 2\pi h^3$ which can be
interpreted as the volume fraction of metal (in a unit cell of the
chain lattice). For an LPC of prolate spheroids, this factor is equal
to $2ab^2/3h^3$. However, for an LPC made of oblate spheroids, the
factor is $2a^2b/3h^3$. It obtains that the propagation distance in
oblate spheroid LPCs is effectively increased by the factor of $a/b>1$
compared to prolate spheroid LPCs. Thus constructing an LPC from thin
nano-disks whose axis of symmetry is parallel to the chain axis may be
beneficial.

\section{Modeling of Wave Packet Propagation}
\label{sec:packets}

\begin{figure}
\centerline{\psfig{file=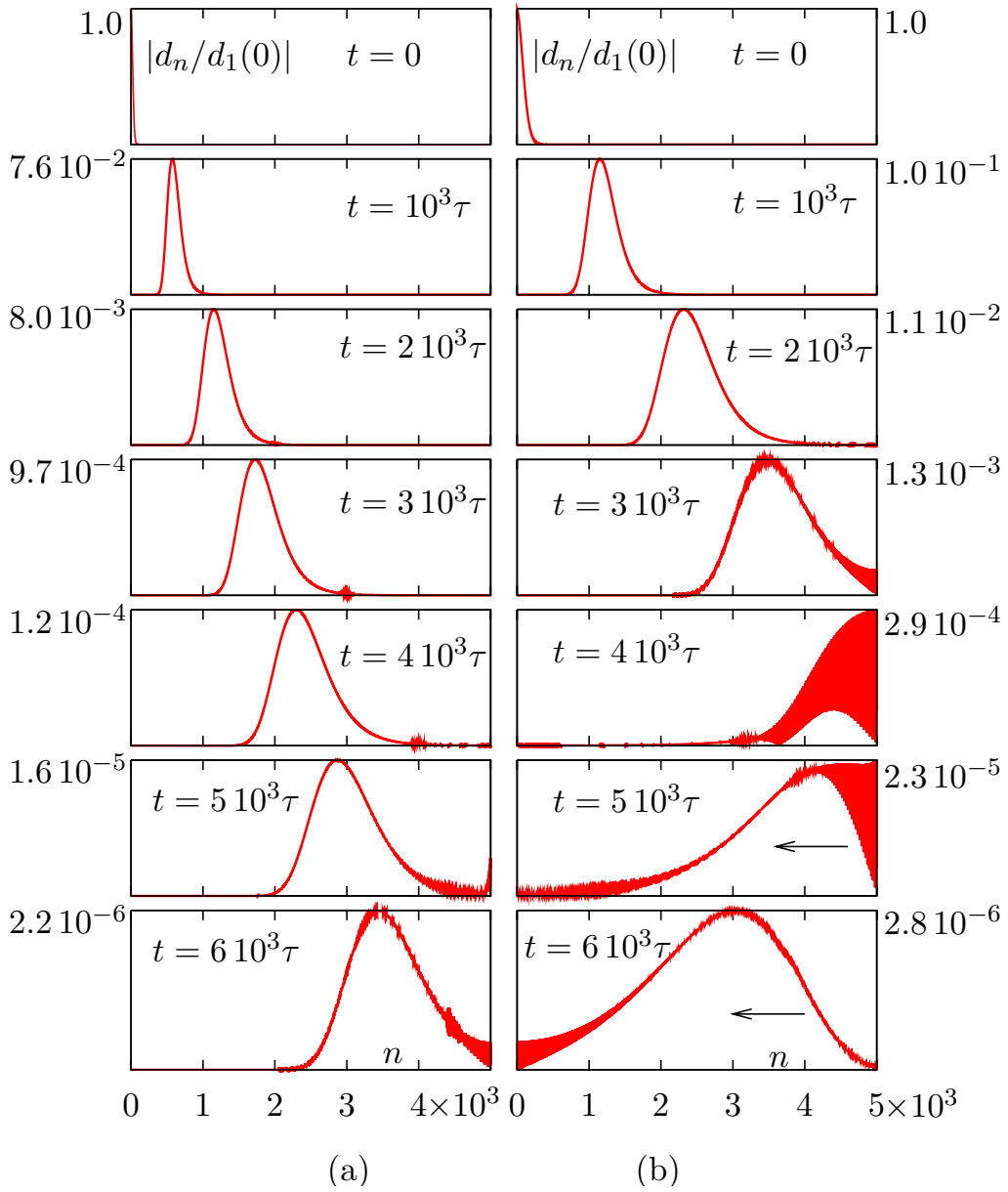,width=14cm,bbllx=140,bblly=380,bburx=470,bbury=760,clip=}}
\caption{\label{fig:packet_ort}
  (color online) Envelopes of transversely polarized wave packets in a
  chain of $N=5000$ prolate nanospheroids with the aspect ratio
  $b/a=0.15$ at different moments of time $t$.  Spheroids are oriented
  so that their axes of symmetry are perpendicular to the chain axis.
  Time is measured in the units of $\tau=h/c_h$.  Column (a):
  $\omega_0=0.1\omega_p$, $v_g\approx 0.57c_h$. Column (b):
  $\omega_0=0.05\omega_p$, $v_g\approx 1.16c_h$.  Arrows indicate that
  the wave packet propagates from right to left after being reflected
  from the far end of the chain.}
\end{figure}

We now demonstrate that superluminal wave packets can indeed propagate
in chains with sufficiently small aspect ratios $b/a$. To this end, we
consider a finite chain of $N$ nanoparticles excited by a pulse with
Gaussian temporal profile incident on the first particle in the chain.
In the time domain, the pulse is described by the formula

\begin{equation}
\label{E_inc_t}
E_n^{\rm ext}(t) = \delta_{n1} {\mathcal E} \exp\left[ -i\omega_0 t - (t/\Delta t)^2 \right] \ ,
\end{equation}

\noindent
where ${\mathcal E}$ is an arbitrary amplitude and $\Delta t$ is the pulse
duration. This function has the Fourier transform

\begin{equation}
\label{E_inc_omega}
\tilde{E}_n^{\rm ext}(\omega) = \delta_{n1} \sqrt{\pi} \Delta t {\mathcal E} \exp\left[ -
  \frac{(\omega - \omega_0)^2}{(\Delta\omega)^2} \right] \ , \ \
\Delta\omega = \frac{2}{\Delta t} \ .
\end{equation}

\noindent
The numerical procedure is as follows. The above expression for
$\tilde{E}_n^{\rm ext}(\omega)$ is used as the free term in the
right-hand side of Eq.~\eqref{CDE}. The equation is solved numerically
by direct matrix inversion for multiple values of $\omega$ sampled in
a sufficiently large interval and with sufficiently small step to
ensure convergence.  This yields a family of numerical solutions
$\tilde{d}_n(\omega)$. The real time quantities $d_n(t)$ are then
obtained by the inverse Fourier transform according to

\begin{equation}
\label{d_n_t}
d_n(t) = \int \tilde{d}_n(\omega) \exp(-i\omega t) \frac{d\omega}{2\pi} \ .
\end{equation}

\noindent
Numerically, this integral is evaluated by the trapezoidal rule.

\begin{figure}
\centerline{\psfig{file=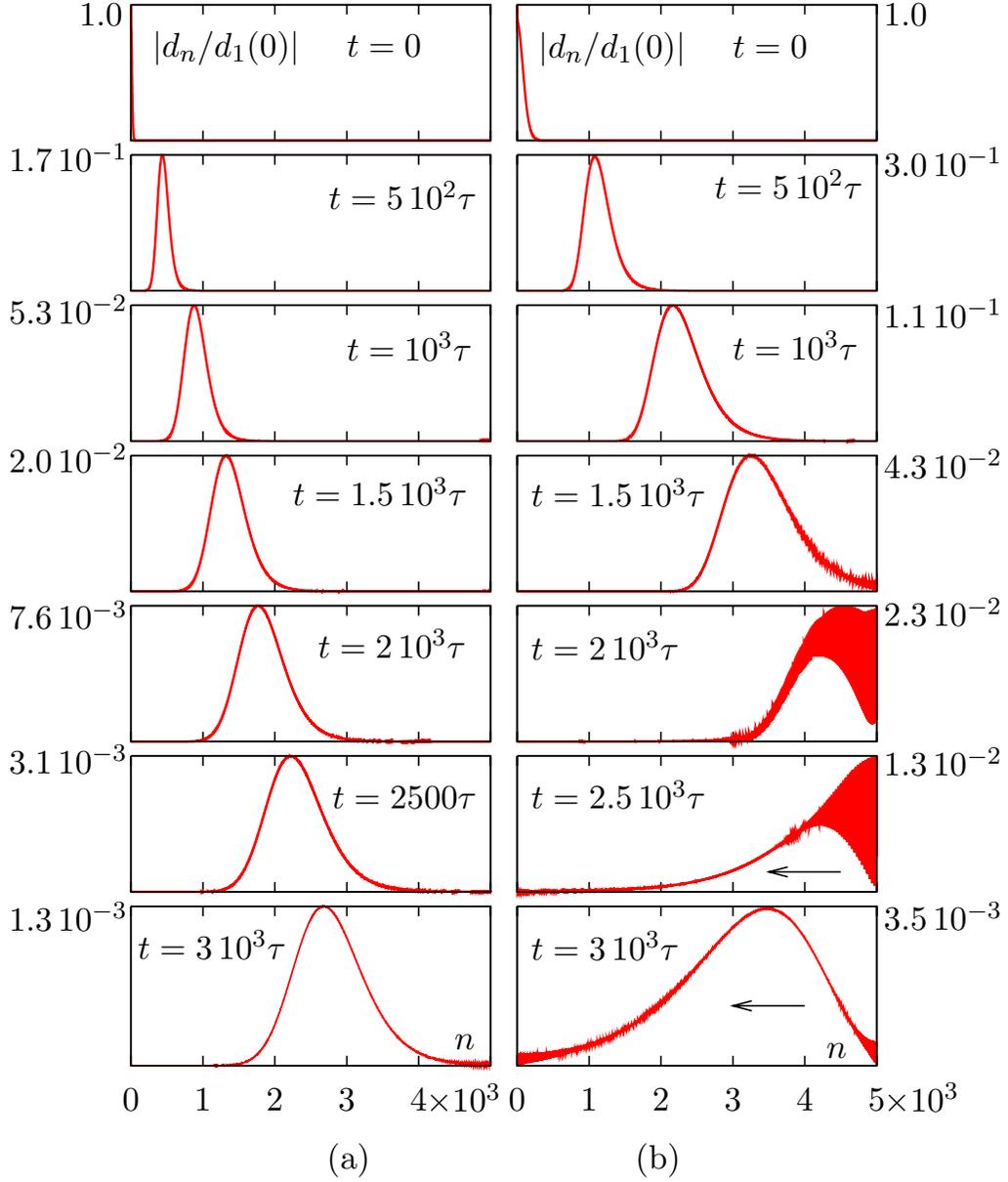,width=14cm,bbllx=140,bblly=380,bburx=470,bbury=760,clip=}}
\caption{\label{fig:packet_par}
  Envelopes of transversely polarized wave packets in a chain of
  $N=5000$ oblate nanospheroids with the aspect ratio $b/a=0.25$ at
  different moments of time $t$. Spheroids are oriented so that their
  axes of symmetry coincide with the chain axis. Time is measured in
  the units of $\tau=h/c_h$. Note that the largest time shown in this
  figure is twice smaller than the respective quantity in
  Fig.~\ref{fig:packet_ort}. Column (a): $\omega_0=0.15\omega_p$,
  $v_g\approx 0.77c_h$. Column (b): $\omega_0=0.05\omega_p$,
  $v_g\approx 2.57c_h$. Arrows indicate that the wave packet
  propagates from right to left after being reflected from the far end
  of the chain.}
\end{figure}

We have simulated transversely polarized wave packets in LPCs of
prolate spheroids with the aspect ratio $b/a=0.15$. Longitudinally
polarized SPPs were simulated in chains of oblate spheroids with the
aspect ratio $b/a=0.25$.  Simulations were performed in a chain
consisting of $N=5\cdot 10^3$ spheroids; the overall length of the
chain was (given $h=25{\rm nm}$) $L=125\mu{\rm m}$. All parameters
were the same as those used for calculating dispersion curves shown in
Figs.~\ref{fig:disp_prol_ort}, with the only exception that Ohmic
losses in the metal were taken into account by means of using the
nonzero Drude relaxation constant $\gamma = 0.002\omega_p$.  Four sets
of simulations have been performed, the first two for the transverse
and the other two for the longitudinal polarization.

In the case of the transverse polarization, two different central
frequencies of the pulse have been used.  The first pulse had the
central frequency $\omega_0=0.1\omega_p$ (correspondingly,
$k_0h/\pi=0.06$ where $k_0=\omega_0/c_h$) and the pulse spectral width
was $\Delta\omega = \omega_0/5$. Thus the excitation was relatively
broad-band but very narrow in the time domain: given the experimental
value of $\omega_p$ for silver, the above spectral width corresponds
to $\Delta t \approx 7.2{\rm fsec}$.  The second pulse had the central
frequency twice smaller than the first, $\omega_0=0.05\omega_p$, with
the same relative spectral width $\Delta\omega = \omega_0/5$. In the
time domain, this corresponds to $\Delta t \approx 14.2{\rm fsec}$.
The central frequencies of the two pulses are shown in
Fig.~\ref{fig:disp_prol_ort}(a) by the horizontal arrows. Amplitudes
of the obtained wave packets are illustrated in
Fig.~\ref{fig:packet_ort} at different moments of time measured in the
units of $\tau=h/c_h$. The maximum time shown on the plots is
$t=6\,10^3\tau \approx 800{\rm fsec}$.

It can be seen that the wave packet with $\omega_0=0.1\omega_p$
propagates away from the source with the subluminal group velocity
$v_g\approx 0.57c_h$. However, the wave packet with the smaller
central frequency propagates at the speed $v_g\approx 1.16c_h$.  These
group velocities are in quantitative agreement with the data shown in
Fig.~\ref{fig:disp_prol_ort}(b). Note that the group velocities can be
evaluated as the slopes of the dispersion curve drawn for $b/a=0.15$
in Fig.~\ref{fig:disp_prol_ort}(a) at the central frequencies
indicated by the horizontal arrows. As expected, the superluminal wave
packet is spreading faster than the subluminal wave packet. This is so
because of the larger value of the second derivative
$\partial^2\omega/\partial^2 q$ at the smaller central frequency. Yet,
near the end of the chain, the duration of the superluminal pulse is
still only $\approx 1{\rm psec}$.

The two sets of simulations for the longitudinal polarization are
shown in Fig.~\ref{fig:packet_par}. Here we have used a chain of
oblate spheroids with the aspect ratio $b/a=0.25$. The central
frequencies of the two pulses were $\omega_0 = 0.25\omega_p$ and
$\omega_0 = 0.1\omega_p$. The pulses' relative spectral widths were
the same as in the case of the transverse polarization, namely, the
pulse spectral width was $\Delta\omega = \omega_0/5$. By tracing the
maximum of each wave packet, we deduce $v_g=0.88c_h$ for
$\omega_0=0.25\omega_p$ and $v_g=2.17c_h$ for $\omega_0=0.1\omega_p$.
This is in full agreement with the group velocities shown in
Fig.~\ref{fig:disp_obl_par}(b).

Finally, note that the decay lengths in each simulation can not be
easily deduced from the time evolution of the maxima of the wave
packets. This is because the propagation is accompanied by both decay
and spreading. The latter takes place even in the absence of Ohmic
losses.

\section{Concluding Remarks}
\label{sec:disc}

In this paper, we have employed the coupled dipole approximation to
compute the dispersion curves and to model propagation of wave packets
of surface plasmon polaritons (SPPs) in linear periodic chains (LPCs)
of metallic nanospheroids. The main novel element of this study, as
compared to the previous work on the
subject~\cite{weber_04_1,simovski_05_1,koenderink_06_1,fung_07_1,park_04_1,citrin_05_1,citrin_06_1,markel_07_2},
is the account of particle nonsphericity.

We have shown that the group velocity, decay length and the bandwidth
of surface plasmon polaritons (SPPs) propagating in linear periodic
chains (LPCs) of metallic nanoparticles can be effectively tuned. The
tunability is achieved by means of varying the nanoparticles aspect
ratio. The decay length can be dramatically increased for Bloch wave
numbers $q$ near the edges of gaps that appear in the first Brillouin
zone of the lattice for sufficiently small aspect ratios. At the same
time, the SPP group velocity is increased up to superluminal values.
By replacing the host medium with vacuum, it is also possible to
excite a wave packet whose group velocity is larger than the speed of
light in vacuum. Such wave packets exist in nature and were observed
experimentally~\cite{wang_00_1,gehring_06_1}.

Comparison of Figs.~\ref{fig:disp_prol_ort} through
\ref{fig:disp_obl_par} reveals that the parameters of two different
LPCs can be tuned so that one LPC supports transversely polarized SPPs
and the other chain supports longitudinally polarized SPPs with {\em
  the same electromagnetic frequency}. This fact can be utilized for
guiding the SPPs through corners (ninety-degree turns in an LPC)
and/or for splitting and coupling the SPPs at T-junctions. Another
potentially interesting element of an integrated photonic circuit a
two-segment straight LPC.  Assume that, at a given frequency, one
segment can support only transversely-polarized SPPs while the other
segment supports only longitudinally polarized SPPs. At the junction
of the two segments, an externally-manipulated (i.e., by magnetic
field) coupling nanospheroid is placed. When the coupling nanospheroid
makes the angle of either $0$ or $\pi$ with respect to the chain axis,
the two segments are decoupled and do not allow direct transmission of
light pulses.  However, if the coupling spheroid is rotated by the
angle of $\pi/4$ with respect to the axis, the transversely-polarized
SPP propagating in the first segment is coupled to the
longitudinally-polarized SPP in the second segment and transmission
along the chain becomes possible. Detailed investigation of these
possibilities will be the subject of future work.

In the case of transverse SPP polarization, the group and phase
velocities of SPPs can be antiparallel. We, however, have found that
the negative group velocity {\em per se} (defined here by the
condition $v_g v_p < 0$) does not necessarily imply superluminal
propagation or a negative time delay as was suggested
previously~\cite{bolda_94_1,dogariu_01_1}. For example, the wave
packet shown in Fig.~\ref{fig:packet_ort}(a) propagates slower than
$c_h$ even though it is composed of waves whose frequencies are in the
negative dispersion region. It is important to realize that the
effects theoretically described in these two references, as well as
the corresponding experimental
observations~\cite{wang_00_1,gehring_06_1}, involve propagation of an
optical pulse from a medium with normal dispersion to a medium with
negative dispersion and the presence of the interface is essential. In
this paper, we are looking at a somewhat different physical situation
when the optical pulse is injected into a waveguide by a predetermined
external source which is located in the near field of the waveguide.
We then observe that the pulse propagates away from the source with
the velocity $\vert v_g \vert$, irrespectively of the sign of the
product $v_g v_p$. Thus the superluminal propagation is obtained when
$\vert v_g \vert > c_h$ but not necessarily when $v_g v_p <0$. 

Antiparallel phase and group velocities that we have observed in the
case of transverse SPP polarization deserve a separate discussion. We
believe that this phenomenon can not be interpreted as ``negative
refraction''. The reason is that the LPCs considered in this paper are
essentially discrete objects and can not be described by effective
medium parameters. The elementary excitations that propagate in LPCs
are Bloch waves rather than sinusoidal waves, and this fact should not
be disregarded.  The region of negative dispersion shown in
Fig.~\ref{fig:disp_prol_ort}(a) starts at $qh \approx 0.2\pi \approx
0.6$.  In general, the chain can be viewed as continuous only when $qh
\ll 1$. The above condition is not satisfied in the region of negative
dispersion. It is, however, not clear {\em a priori}, how strong this
inequality should be for the effective medium approximation to be
valid. In the specific case of LPCs, one can consider the following
argument. We expect the effective medium parameters such as the
permittivity $\epsilon(\omega)$ or the refraction index $n(\omega)$ to
be single-valued functions of their argument. However, for every point
on the negative slope section of the dispersion curves shown in
Fig.~\ref{fig:disp_prol_ort}(a), there is another point on the same
curve with the same frequency but a smaller value of $q$. This second
point is located on the linear, small-$q$ section of the dispersion
curve.  Although this small-$q$ section is difficult to find
numerically (and, as a result, is often overlooked), it exists. It is
therefore logical to assume that, if a chain be assigned some
effective medium parameter for a given frequency $\omega$, this
parameter must be computed using the point on the small-$q$ section of
the dispersion curve. The latter exhibits positive (and linear)
dispersion. In the above argument, we have disregarded the possibility
of introducing non-local effective medium parameters which are
characteristic, for example, of chiral media and can result in
negative dispersion~\cite{agranovich_06_1}.  However, the physical
object that we consider in this paper is essentially non-chiral.

The authors can be reached at: \\
{\tt algov@seas.upenn.edu} and {\tt vmarkel@mail.med.upenn.edu}.

\bibliographystyle{prsty} 
\bibliography{abbrev,master,book}
\end{document}